\documentclass[fleqn,10pt]{wlscirep}
\usepackage[utf8]{inputenc}
\usepackage[T1]{fontenc}
\usepackage{braket}
\usepackage{physics}
\usepackage{amsmath}
\usepackage{graphics}
\usepackage{caption}
\usepackage{subcaption}
\usepackage{verbatim}

\title{The quantum Zeno and anti-Zeno effects in the strong coupling regime}

\author[1, +]{Ghazi Khan}
\author[1, +]{Hudaiba Soomro}
\author[1, +]{Muhammad Usman Baig}
\author[2]{Irfan Javed}
\author[1,*]{Adam Zaman Chaudhry}
\affil[1]{School of Science \& Engineering, Lahore University of Management Sciences (LUMS), Opposite Sector U, D.H.A, Lahore 54792, Pakistan}
\affil[2]{University of New Brunswick (UNB), Department of Mathematics and Statistics, Frederiction, Canada}

\affil[*]{adam.zaman@lums.edu.pk}

\affil[+]{These authors contributed equally to this work.}

\begin{abstract}
It is well known that repeated projective measurements can either speed up (the Zeno effect) or slow down (the anti-Zeno effect) quantum evolution. Until now, however, studies of these effects for a two-level system interacting strongly with its environment have focused on repeatedly preparing the excited state of the two-level system via the projective measurements. In this paper, we consider the repeated preparation of an arbitrary state of a two-level system that is interacting strongly with an environment of harmonic oscillators. To handle the strong interaction, we perform a polaron transformation, and thereafter use a perturbative approach to calculate the decay rates for the system. Upon calculating the decay rates, we discover that there is a transition in their qualitative behaviors as the state being repeatedly prepared moves away from the excited state towards a superposition of the ground and excited states. Our results should be useful for the quantum control of a two-level system interacting with its environment.
\end{abstract}
\begin{document}
\flushbottom
\maketitle
%
%
\thispagestyle{empty}
\section*{Introduction}

By subjecting a quantum system to frequent and repeated projective measurements, we can slow down its temporal evolution, an effect referred to as the quantum Zeno effect (QZE) \cite{Sudarshan1977, FacchiPhysLettA2000, FacchiPRL2002, FacchiJPA2008, WangPRA2008, ManiscalcoPRL2008, FacchiJPA2010, MilitelloPRA2011, RaimondPRA2012, SmerziPRL2012, WangPRL2013, McCuskerPRL2013, StannigelPRL2014, ZhuPRL2014, SchafferNatCommun2014, SignolesNaturePhysics2014, DebierrePRA2015, AlexanderPRA2015, QiuSciRep2015, HePRA2018, HanggiNJP2018, HePRA2019, MullerAnnPhys}. Contrary to this effect is the quantum anti-Zeno effect (QAZE), via which the temporal evolution of the system is accelerated due to repeated projective measurements separated by relatively longer measurement intervals \cite{KurizkiNature2000, RaizenPRL2001, BaronePRL2004, KoshinoPhysRep2005, BennettPRB2010, YamamotoPRA2010, ChaudhryPRA2014zeno, Chaudhryscirep2017b, HePRA2017, WuPRA2017, Chaudhryscirep2018, WuAnnals2018, ChaudhryEJPD2019a}. Both these effects have garnered great interest not only due to their theoretical relevance to quantum foundations but also due to their applications to quantum technologies. For example, the QZE has shown to be a promising resource for quantum computing and quantum error correction \cite{fransonPRA2004, pazsilvaPRA2012}. The QAZE, on the other hand, has interestingly been useful in, say, accelerating chemical reactions, suggesting the possibility of quantum control of a chemical reaction \cite{prezhdopr2000}. 

By and large, studies of the QZE and the QAZE have focused on population decay \cite{KurizkiNature2000, RaizenPRL2001, BaronePRL2004, KoshinoPhysRep2005, ManiscalcoPRL2006, SegalPRA2007, ZhengPRL2008, BennettPRB2010, YamamotoPRA2010, AiPRA2010, ThilagamJMP2010, ThilagamJCP2013} and pure dephasing models \cite{ChaudhryPRA2014zeno}. While the former studies measurements performed on a single two-level system, the latter considers the effect of dephasing on the QZE and the QAZE. A few works have gone beyond these regimes. Ref.~{\renewcommand{\citemid}{}\cite[]{Chaudhryscirep2016}}, for instance, presents a general framework to calculate the effective decay rate for an arbitrary system-environment model in the weak coupling regime and finds it to be the overlap of the spectral density of the environment and a filter function that depends on the system-environment model, the measurement interval, and the measurement being performed. This approach, however, fails in the strong coupling regime where perturbation theory cannot be applied in a straightforward manner \cite{ChaudhryPRA2014zeno}. For a single two-level system coupled strongly to an environment of harmonic oscillators, Ref.~{\renewcommand{\citemid}{}\cite[]{Chaudhryscirep2017a}} makes the problem tractable by going to the polaron frame and finding that for the excited state, the decay rate very surprisingly decreases with an increase in the system-environment coupling strength. This effect is further investigated in Ref.~\renewcommand{\citemid}{}\cite[]{irfan}, which studies a two-level system coupled simultaneously to a weakly interacting dissipative-type environment and a strongly interacting dephasing-type one. It is found that even in the presence of both types of interactions, the strongly coupled reservoir can inhibit the influence of the weakly coupled reservoir on the central quantum system. 

To date, the role of the state that is repeatedly prepared has been left unexplored, especially in the strong coupling regime.  For example, it remains unanswered whether increasing the coupling strength of a strongly coupled reservoir would lead to the decay rate decreasing for states other than the excited state. This forms the basis of our investigation in this paper. We work out the decay rates for a two-level system, strongly interacting with a bath of harmonic oscillators, that is repeatedly prepared in an arbitrary quantum state. To make the problem tractable, we first go to the polaron frame, where the system-environment coupling is effectively weakened, and thereafter use time-dependent perturbation theory to evolve the system state and find its decay rate. While we reproduce the results presented for an excited state in Ref.~{\renewcommand{\citemid}{}\cite[]{Chaudhryscirep2017a}}, we observe a stark difference when the initial state is chosen to be a superposition of the excited and ground states. To be precise, the qualitative variation of the decay rate with the system-environment coupling gets inverted. To describe these results, we coin the terms ``$z$-type" and ``$x$-type", identifying the behavior displayed by Ref.~{\renewcommand{\citemid}{}\cite[]{Chaudhryscirep2017a}} as the $z$-type behavior while the inverted behavior is termed as the $x$-type behavior. We investigate the transition between these behaviors. These results should be useful in the study of open quantum systems in the strong coupling regime.

\section*{Results}
\subsection*{Effective decay rate for strong system-environment coupling}

\begin{figure}
\centering
\begin{subfigure}{.5\textwidth}
  \centering
  \includegraphics[width=\textwidth]{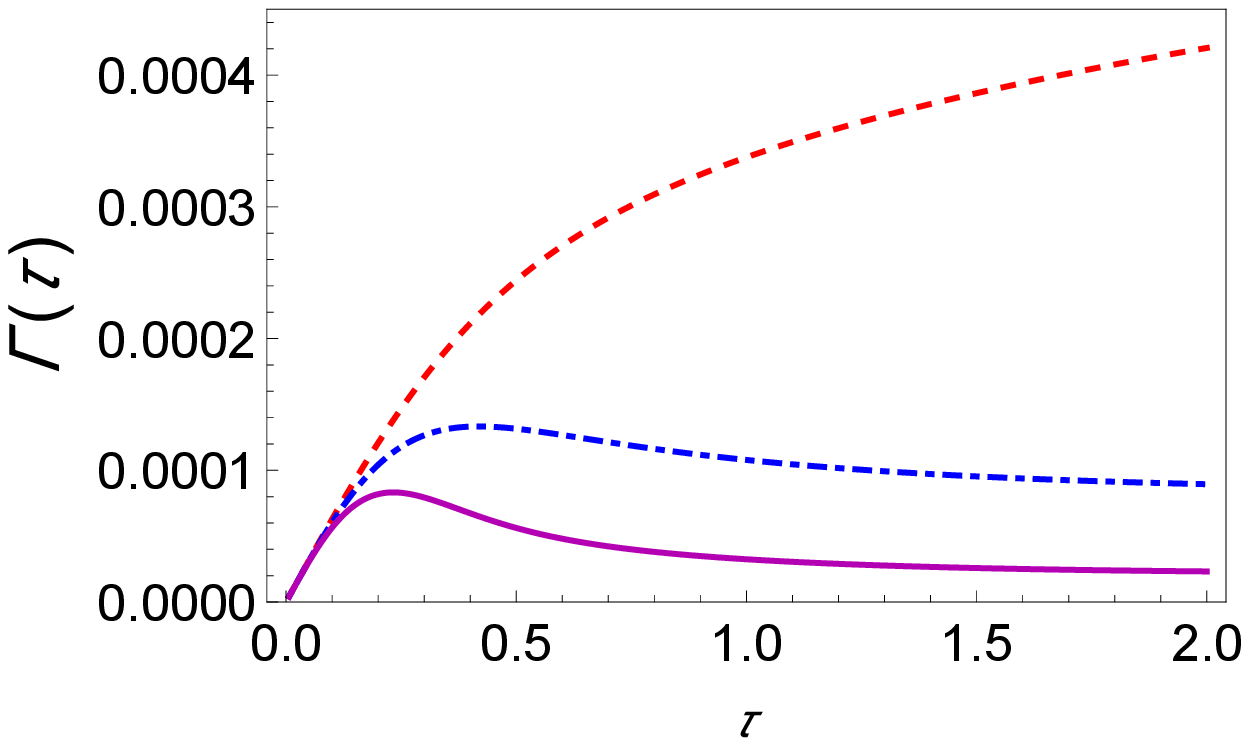}
  \caption{}
  \label{fig1a}
\end{subfigure}%
\begin{subfigure}{.5\textwidth}
  \centering
  \includegraphics[width=\textwidth]{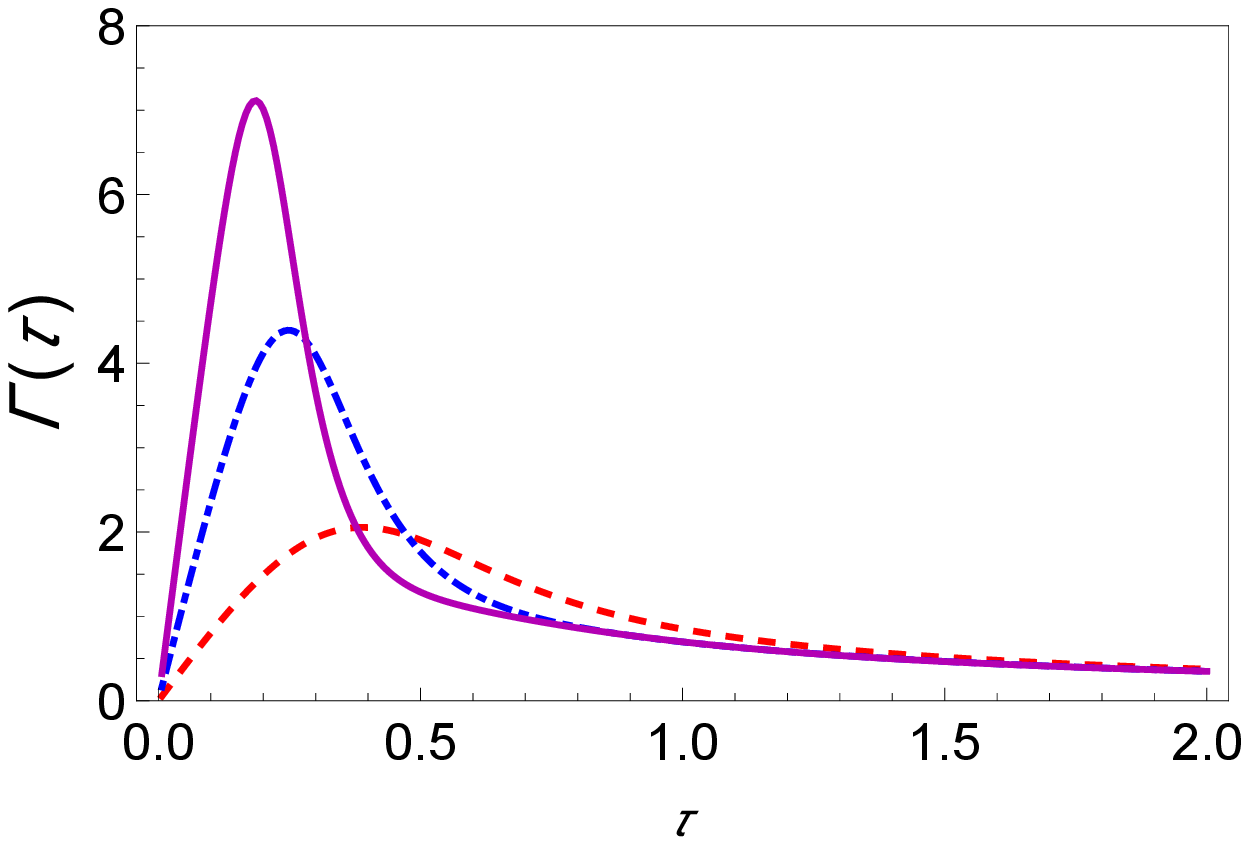}
  \caption{}
  \label{fig1b}
\end{subfigure}
\caption{\textbf{Variation of the effective decay rate with change in coupling strength.} \textbf{(a)} Graph of $\Gamma(\tau)$ (at zero temperature) for the initial state $\ket{0}$ regime as a function of $\tau$ with $G=1$ (red, dashed), $G=2$ (dot-dashed, blue), and $G=3$ (solid magenta curve). We have used a super-Ohmic environment ($s=2$), with $\omega_c = 1$, $\epsilon=1$, and $\Delta = 0.05$. \textbf{(b)} Behavior of $\Gamma(\tau)$ (at zero temperature) for the initial state $(\ket{0}+\ket{1})/\sqrt{2}$ as a function of $\tau$ with $G=1$ (red, dashed), $G=2$ (dot-dashed, blue), and $G=3$ (solid magenta curve). We have used a super-Ohmic environment ($s=2$), with $\omega_c = 1$, $\epsilon=1$, and $\Delta = 0.05$.}
\label{fig1}
\end{figure}

We start from the the paradigmatic spin-boson model \cite{LeggettRMP1987, Weissbook, Breuernonmarkovianreview} with the system-environment Hamiltonian written as (we work in dimensionless units with $\hbar = 1$ throughout) 
\begin{equation}
    H_{L} = \frac{\epsilon}{2}\sigma_z + \frac{\Delta}{2} \sigma_x + \sum_{k}\omega_k b_k^{\dagger}b_k  + \sigma_z(\sum_{k}g_k^{\ast} b_k^{\dagger} + g_k b_k),
\label{eq1}
\end{equation}
where $H_{S,L} = \frac{\epsilon}{2}\sigma_z + \frac{\Delta}{2} \sigma_x$ is the system Hamiltonian, $H_B = \sum_{k=1}\omega_k b_k^{\dagger}b_k$ is the environment Hamiltonian, and $V_L = \sigma_z(\sum_{k}g_k^{\ast} b_k^{\dagger} + g_k b_k)$ is the system-environment coupling. Note that $L$ denotes the lab frame, $\epsilon$ is the energy splitting of the two-level system, $\Delta$ is the tunneling amplitude, and the $\omega_k$ are frequencies of the harmonic
oscillators in the harmonic oscillator environment interacting with the system. The creation and annihilation operators of these oscillators are represented by the $b_k^\dagger$ and $b_k$ operators respectively. In the strong interacting regime, we cannot treat the interaction perturbatively. Moreover, the initial system-environment correlations are significant and thus cannot be neglected to write the initial state as a simple product state \cite{ChaudhryPRA2013a, ChaudhryPRA2013b}. To make the problem tractable, we perform a polaron transformation \cite{SilbeyJCP1984, VorrathPRL2005, jang2008theory, ChinPRL2011, lee2012accuracy, LeePRE2012, gelbwaser2015strongly}, which yields an effective interaction that is weak. More precisely, the transformation to the polaron frame is given by $H = U_{P}H_{L}U_P^{\dagger}$, where $U_P= e^{-\frac{\chi}{2}\sigma_z}$ and $\chi= \sum_k(\frac{g_k}{\omega_k}b_k - \frac{g_k^{\ast}}{\omega_k}b_k^{\dagger})$. We then get the transformed Hamiltonian
\begin{equation}
    H= \frac{\epsilon}{2}\sigma_z + \sum_{k}\omega_k b_k^{\dagger}b_k+
    \frac{\Delta}{2}(\sigma_{+}e^{\chi} + \sigma_{-}e^{-\chi}).
    \label{eq2}
\end{equation}
For future convenience, we define $H_0 = \frac{\epsilon}{2}\sigma_z + \sum_{k}\omega_k b_k^{\dagger}b_k$. Now, if $\Delta$ is taken as being small, the system and environment interact effectively weakly in the polaron frame despite interacting strongly in the lab frame. Let $\ket{0}$ represent the excited state  of our two-level system, and $\ket{1}$ be its ground state. Then, writing an arbitrary initial state of the two-level system as $\ket{\psi}= \zeta_{1} \ket{0} + \zeta_{2} \ket{1}$ with $\zeta_1 = \cos{(\theta/2)}$ and $\zeta_2 = e^{i\phi }\sin{(\theta/2)}$, we find the time-evolved density matrix by means of time-dependent perturbation theory. It is important to note that while the initial system-environment state cannot simply be taken as a simple uncorrelated product state in the `lab' frame \cite{ChaudhryPRA2013a, ChaudhryPRA2013b}, we can do so in the polaron frame since the system and its environment are interacting weakly in the polaron frame. We subsequently perform repeated measurements after time intervals of duration $\tau$ to see if the system state is still $\ket{\psi}$. The survival probability at time $\tau$ is $s(\tau)= \Tr_{S,B}\{{P_{\psi}\rho(\tau)}\}$, where $\rho(\tau)$ is the combined density matrix of the system and the environment at time $t=\tau$ in the polaron frame before the projective measurement while $P_{\psi}= U_P\ketbra{\psi}{\psi}U_P^{\dagger}$. This survival probability is then 
\begin{equation}
    s(\tau)=\Tr_{S,B}\{P_{\psi}e^{-iH\tau} P_{\psi}\frac{e^{-\beta H_{0}}}{Z} P_{\psi}e^{iH\tau}\},
\label{eq3}
\end{equation}
with $Z$ a normalization factor. The subsequent detailed calculation is in the Methods section. For the most general case, this yields a rather extensive expression. In this section, however, we present the expressions for some important cases only. First, let us consider the initial state $\ket{0}$, that is $\ket{\psi}$ with $\zeta_{1}=1$ and $\zeta_{2}=0$. It is found that
\begin{eqnarray}
    s(\tau) = 1-2\Re{\frac{\Delta^2}{4}\int_{0}^{\tau} dt_{1}\int_{0}^{t_{1}}dt_{2}\abs{\zeta_{1}}^2 e^{-i\epsilon t_1} e^{i\epsilon t_2}C(t_{2}-t_{1})},
    \label{eq4}
\end{eqnarray}
where $C(t_{2}-t_{1})$ is the environment correlation function and is given by $C(t_{2}-t_{1})= e^{-\Phi_{C}^{\ast}(t_2 - t_1)}$, where $\Phi_{C}(\tau)= \Phi_R(\tau) - i\Phi_I(\tau)$ with $\Phi _{R}= 4\int_{0}^{\infty}d\omega J(\omega)\frac{1-\cos{\omega t}}{\omega^2}\coth({\frac{\beta\omega}{2}})$ and $\Phi _{I}=4\int_{0}^{\infty}d\omega J(\omega)\frac{\sin{\omega t}}{\omega^2}$. The environment spectral density $J(\omega)$ has been introduced as $\sum_{k}\abs{g_k}^2(\cdots) \rightarrow \int_{0}^{\infty}d\omega J(\omega)(\cdots)$. Since the system-environment coupling in
the polaron frame is weak, we can neglect the accumulation of correlations between the system and the environment and write the survival probability at time $t=N\tau$, or $s(t=N\tau)$, as $[s(t)]^N$, where $N$ denotes the number of measurements performed after time $t=0$. Now, we may write $s(t=N\tau)\equiv e^{-\Gamma(\tau)N\tau}$ to define an effective decay rate $\Gamma(\tau)$ for our quantum state. It follows that $\Gamma(\tau)= -\frac{1}{\tau}[\ln{s(\tau)]}$. Expanding $\ln[s(\tau)]$ up to the second order in $\Delta$, we work out the decay rate to $\frac{1-s(\tau)}{\tau}$. Furthermore, in order to numerically investigate how the decay rate varies with the measurement interval $\tau$, we model the spectral density as $J(\omega)= G\omega^{s}\omega_{c}^{1-s}e^{-\omega/\omega_c}$, where $G$ is a dimensionless parameter characterizing the strength of the system-environment coupling, $\omega_c$ is the cut-off frequency, and $s$ is the so-called Ohmicity parameter. Throughout, we present results for a super-Ohmic environment with $s=2$. For this case, we obtain $\Phi _{R}=G(4-\frac{4}{1+\omega_c^2 t^2})$ and $\Phi _{I}= \frac{4Gt}{(\omega_c(\frac{1}{\omega_c^2}+t^2))}$. For simplicity, we also choose to work at zero temperature. We then obtain
\begin{equation}
    \Gamma(\tau)= \frac{\Delta^2}{2\tau}\Re{\int_{0}^{\tau} dt_{1}\int_{0}^{t_{1}}dt_{2}e^{-i\epsilon (t_1-t_2)} e^{-G(4-\frac{4}{1+\omega_c^2 (t_2-t_1)^2})}e^{-i\frac{4G(t_2-t_1)}{\omega_c(\frac{1}{\omega_c^2}+(t_2-t_1)^2)}}}.
    \label{eq5}
\end{equation}
Working out the double integrals numerically then, we plot Eq.~(\ref{eq5}) in Fig.~(\ref{fig1a}) for various system-environment coupling strengths, thereby exactly reproducing the qualitative behavior presented by Ref.~{\renewcommand{\citemid}{}\cite[]{Chaudhryscirep2017a}} for this case. The behavior of $\Gamma(\tau)$ as a function of $\tau$ allows us to identify the Zeno and anti-Zeno regimes. The Zeno regime is marked by the region where decreasing $\tau$ leads to a decrease in $\Gamma(\tau)$. The anti-Zeno regime, alternatively, is marked by the region where decreasing $\tau$ leads to an increase in $\Gamma(\tau)$ \cite{KurizkiNature2000, ChaudhryPRA2014zeno, Chaudhryscirep2016, WuAnnals2018, SegalPRA2007, ThilagamJMP2010}. With these criteria, whereas we observe only the QZE for $G=1$ in Fig.~(\ref{fig1a}), we also see the QAZE for $G=2$ and $G=3$. Increasing $G$ bears forth a significant qualitative change in the QZE/QAZE behavior of the central quantum system. Moreover, as is evident, increasing $G$ actually decreases the effective decay rates. 

\begin{figure}[h!]
\centering
\begin{subfigure}{.5\textwidth}
  \centering
  \includegraphics[width=0.95\linewidth]{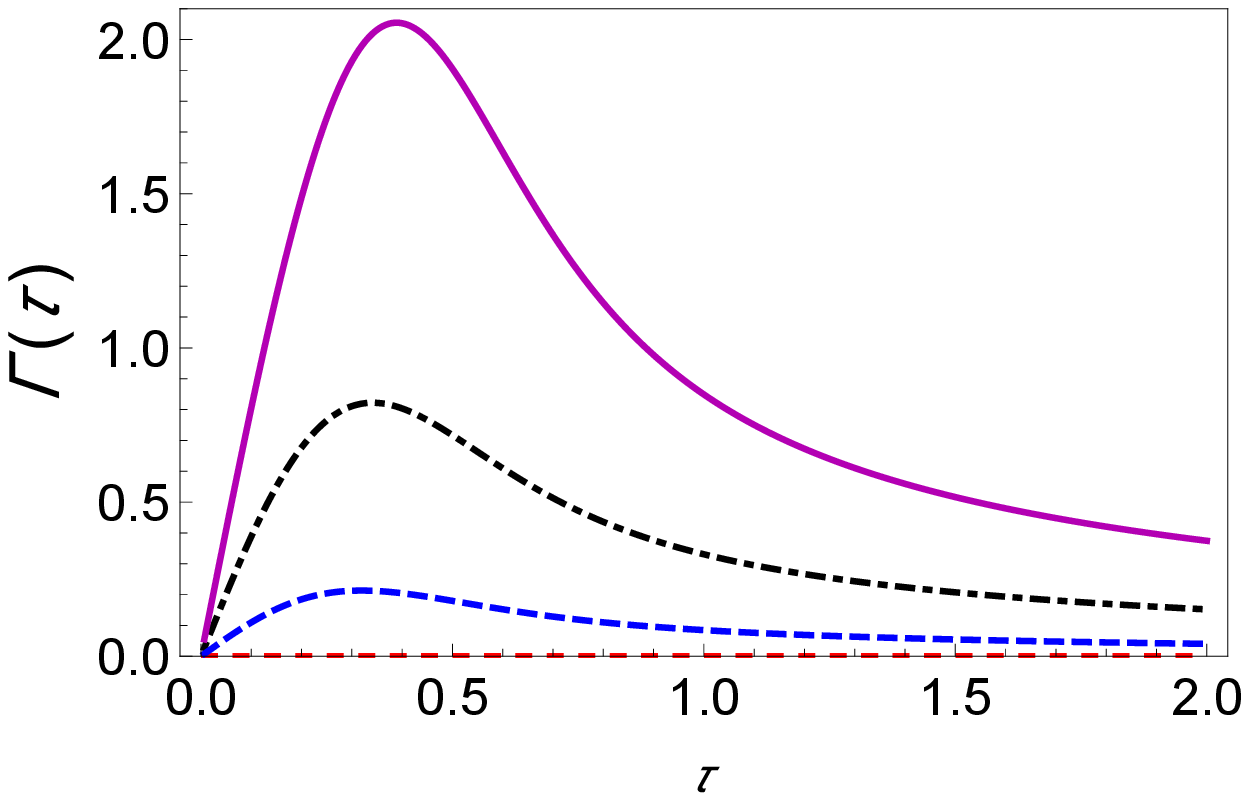}
  \caption{}
  \label{fig2a}
\end{subfigure}%
\begin{subfigure}{.5\textwidth}
  \centering
  \includegraphics[width=0.95\linewidth]{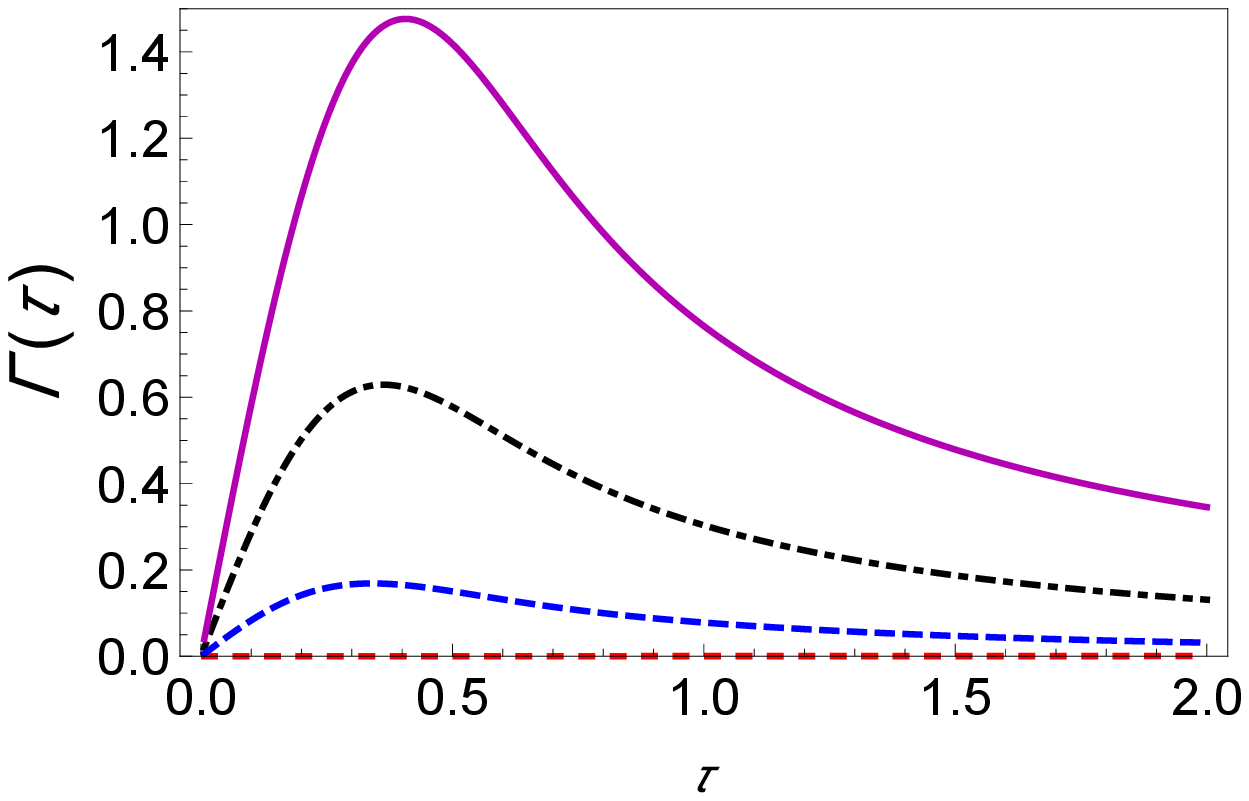}
  \caption{}
  \label{fig2b}
\end{subfigure}
\caption{\textbf{Variation of the effective and modified decay rates with change in the initial state.} \textbf{(a)} The effective decay rate, $\Gamma(\tau)$ (at zero temperature) as a function of $\tau$ for different initial states, $\theta =0$ (dashed, red curve), $\theta =\pi/8$ (dashed, blue curve), $\theta =\pi/4$ (dot-dashed, black curve), and $\theta = \pi/2$ (solid magenta curve). \textbf{(b)} The modified decay rate, $\Gamma^{n}(\tau)$ (at zero temperature) as a function of $\tau$ for different initial states with the same labels as (a). For both cases, we have used a super-Ohmic environment ($s=2$) with $\omega_c = 1$, $\epsilon=1$, and $\Delta= 0.05$.}
\label{fig2}
\end{figure}

Having shown that our generalized approach reproduces the known qualitative behavior for the initial state $\ket{0}$, we move on to considering the superposition state $(\ket{0}+\ket{1})/\sqrt{2}$ as our initial state. That our approach is independent of the initial state chosen allows us to consider this particular superposition state and we choose to investigate it because it happens to be the farthest away from both the ground state and the excited state. Again working at zero temperature, we find the decay rate to be
\begin{equation}
    \begin{aligned}
    \begin{split}
    \Gamma^{(0)}(\tau)=& \frac{1}{\tau}\bigg[1-\frac{1}{4}\bigg[2e^{\beta\epsilon/2}+ 2\Re\bigg\{e^{i\epsilon\tau}e^{\beta\epsilon/2}e^{-\Phi_{C}(\tau)}\bigg\}\\
    &+ 2\Re\bigg\{i\frac{\Delta}{2}\int_{0}^{\tau}dt_1 \bigg(e^{-i\epsilon t_1}e^{\beta\epsilon/2}e^{-\Phi_{C}^{\ast}(t_1)}+e^{i\epsilon t_1}e^{\beta\epsilon/2}e^{-\Phi_{C}(t_1)}+e^{i\epsilon\tau}e^{-i\epsilon t_1}e^{\beta\epsilon/2}e^{-\Phi_{C}^{\ast}(t_1-\tau)}\\
    &+e^{-i\epsilon\tau} e^{i\epsilon t_1}e^{\beta\epsilon/2}e^{-\Phi_{C}^{\ast}(t_1-\tau)}e^{-2i\Phi _{I(\tau)}}e^{2i\Phi _{I(t_1)}}\bigg)\bigg\}\\
    &-2\Re\bigg\{\frac{\Delta^2}{4}\int_{0}^{\tau}dt_1\int_{0}^{t_1}dt_2 \bigg( 
    e^{-i\epsilon t_1}e^{i\epsilon t_2}e^{\beta\epsilon/2}e^{-\Phi_{C}^{\ast}(t_2-t_1)}e^{2i\Phi _{I}(t_2)}e^{-2i\Phi_{I}(t_1)}+e^{i\epsilon t_1}e^{-i\epsilon t_2}e^{\beta\epsilon/2}e^{\Phi_{C}^{\ast}(t_2-t_1)}\\
    &+e^{i\epsilon\tau}e^{-i\epsilon t_1}e^{i\epsilon t_2}e^{\beta\epsilon/2}W' e^{i\Phi_{I}(t_2)}e^{-i\Phi_{I}(t_1)}e^{i\Phi_{I}(t)}+e^{-i\epsilon\tau}e^{i\epsilon t_1}e^{-i\epsilon t_2}e^{\beta\epsilon/2}W' e^{-i\Phi_{I}(t_2)}e^{i\Phi_{I}(t_1)}e^{-i\Phi _{I}(\tau)}\bigg)\bigg\}\\
    &+ \frac{\Delta^2}{4}\int_{0}^{\tau}dt_1\int_{0}^{\tau}dt_2\bigg(e^{i\epsilon t_1}e^{-i\epsilon t_2}e^{\beta\epsilon/2}e^{-\Phi_{C}^{\ast}(t_2-t_1)}+ e^{-i\epsilon t_1}e^{i\epsilon t_2}e^{\beta\epsilon/2}
    e^{-\Phi_{C}^{\ast}(t_2-t_1)}e^{2i\Phi_{I}(t_2)}e^{-2i\Phi_{I}(t_1)}\\
    &+e^{i\epsilon\tau}e^{-i\epsilon t_1}e^{-i\epsilon t_2}e^{\beta\epsilon/2}W e^{-i\Phi_{I}(t_2)}e^{-i\Phi _{I}(t_1)}e^{i\Phi_{I}(\tau)}+e^{-i\epsilon\tau}e^{i\epsilon t_1}e^{i\epsilon t_2}e^{\beta\epsilon/2}We^{i\Phi _{I}(t_2)}e^{i\Phi_{I}(t_1)}e^{i\Phi_{I}(\tau)}
    \bigg)\bigg]\bigg],
    \end{split}
    \end{aligned}
\label{eq6}
\end{equation}
where $W$ and $W'$ comprise further bath correlation terms and are given in the Supplementary Information. As before, we plot this decay rate for different system-environment coupling strengths in Fig.~(\ref{fig1b}), which reveals a marked difference from the qualitative behavior displayed by Fig.~(\ref{fig1a}). As we increase the coupling strength, the decay rate peak rises, which is precisely opposite to what we saw in the previous case. Not only does using the superposition state change the numerical values of the decay rate, but it also essentially inverts the inhibiting effect that an increase in the coupling strength had on the decay rate before. This result makes sense because in the strong coupling regime, the system-environment coupling acts as a protection for its eigenstates, meaning that the eigenstates of the interaction term actually benefit from an increased coupling with the environment in that they remain alive for longer times\cite{Chaudhryscirep2017a}. This protection is, however, lost as we move away from $\ket{0}$ on the Bloch sphere as is apparent in Fig.~(\ref{fig2a}), where we plot the decay rates for varying polar angles.

\begin{figure}
    \centering
    \includegraphics[width=0.5\textwidth]{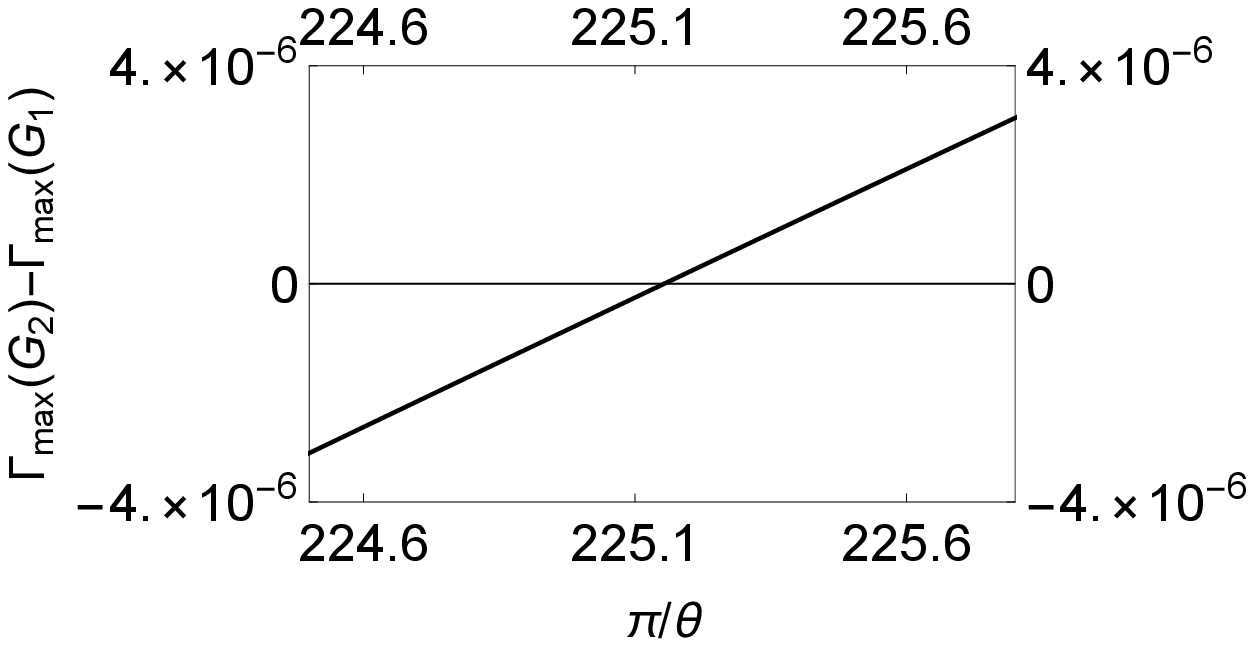}
    \caption{Graph of the difference between the maxima of decay rates corresponding to $G_1$ and $G_2$, $\Gamma_{\text{max}(G_2)}-\Gamma_{\text{max}(G_1)}$ against $\pi/\theta$.}
    \label{fig3}
\end{figure}

\begin{figure}[h!]
     \centering
     \begin{subfigure}[b]{0.33\textwidth}
         \centering
         \includegraphics[width=\textwidth]{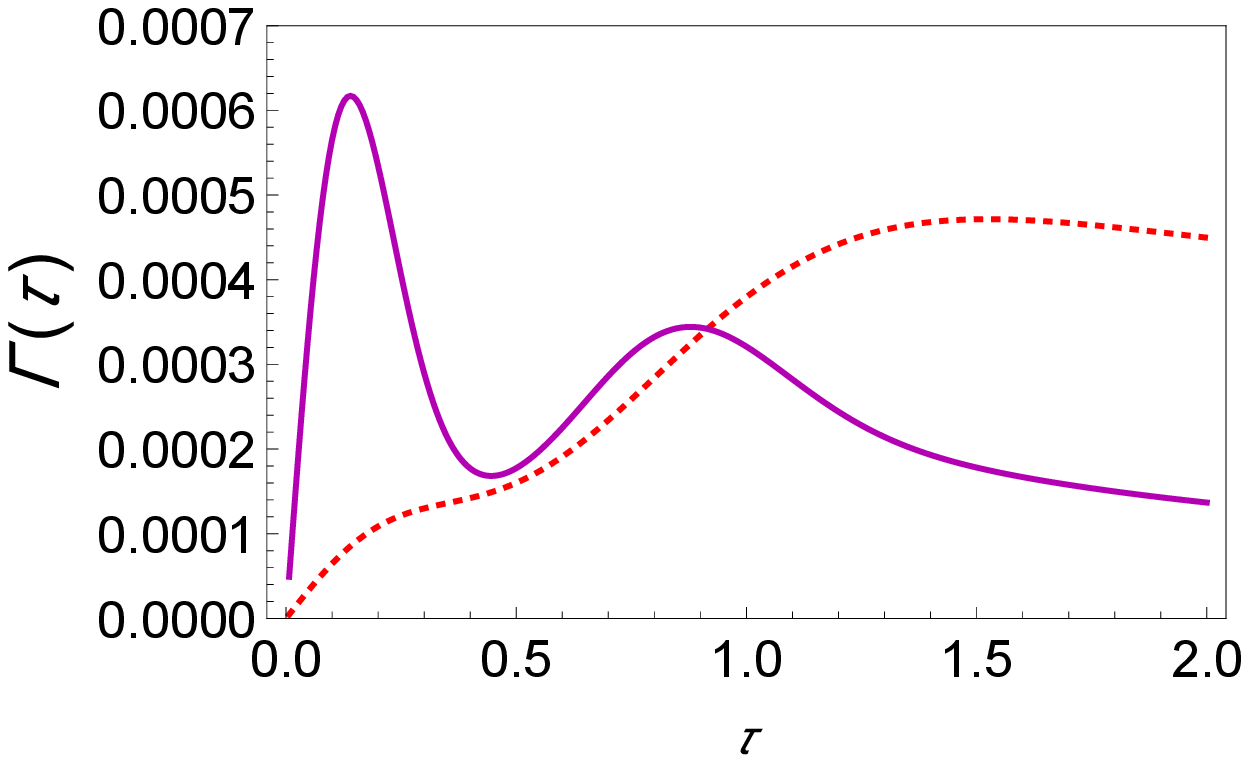}
         \caption{}
         \label{fig4a}
     \end{subfigure}
     \hfill
     \begin{subfigure}[b]{0.33\textwidth}
         \centering
         \includegraphics[width=\textwidth]{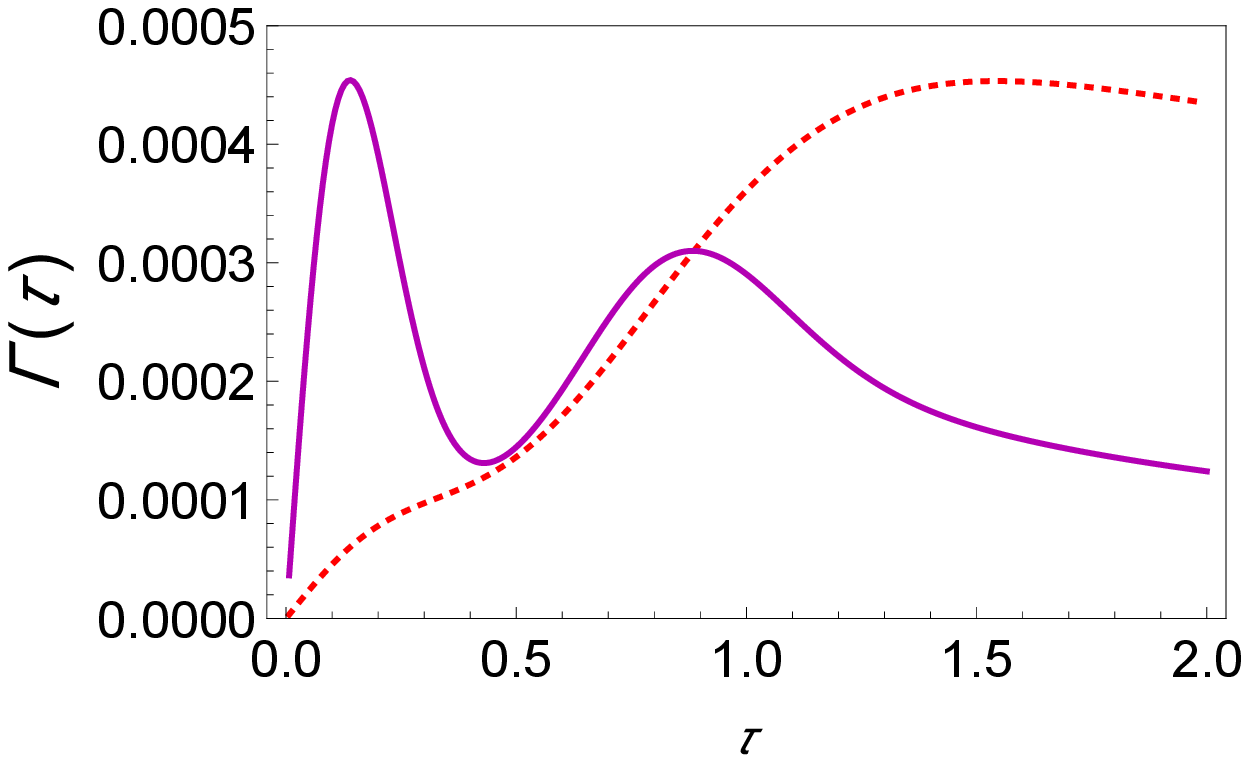}
         \caption{}
         \label{fig4b}
     \end{subfigure}
     \hfill
     \begin{subfigure}[b]{0.33\textwidth}
         \centering
         \includegraphics[width=\textwidth]{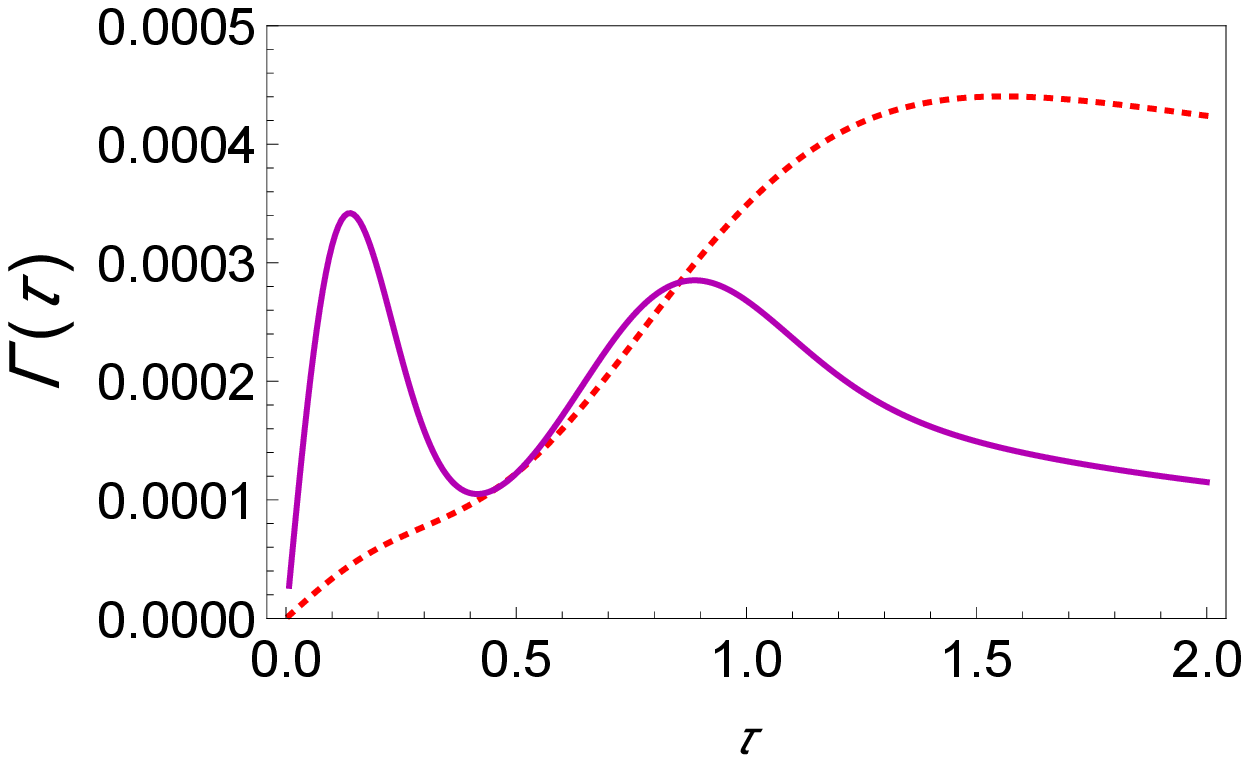}
         \caption{}
         \label{fig4c}
     \end{subfigure}
        \caption{\textbf{Transitory behavior in the effective decay rates}
        \textbf{(a)} Graph of $\Gamma(\tau)$ (at zero temperature) with the initial state corresponding to $\theta=\pi/200$, for $G=1$ (solid magenta curve) and $G=3$ (dashed, red curve). \textbf{(b)} Same as (a) with initial state corresponding to $\theta_c = \pi/225$, showing critical behavior. \textbf{(c)} Same as (a) except $\theta=\pi/250.$}
        \label{fig4}
\end{figure}

If $\Gamma(\tau)$ is plotted against $\tau$ for different values of $\theta$, it is found that for any coupling strength, all the initial states have decay rates with one maximum. If we now assume two different coupling strengths $G_1$ and $G_2$, and we may assume $G_2>G_1$ without loss of generality, we notice that the decay rates exhibit either ``$z$-type" or ``$x$-type" behaviour. For states we term as having $z$-type behaviour, the maximum of $\Gamma(\tau)$ corresponding to $G_1$ is greater as is characteristic of the state $\ket{0}$ in Fig.~(\ref{fig1a}). Similarly, we term states as showing $x$-type behaviour if the maximum of $\Gamma(\tau)$ corresponding to $G_1$ is lesser. Hence, for any $G_1$ and $G_2$, there has to exist a value of the angle $\theta$ at which we see a transition between these two behaviors. To show the existence of this critical value of $\theta$, which we label as $\theta_c$, we plot the difference between the respective maxima of decay rates corresponding to $G_1$ and $G_2$ against $\theta$ (see Fig.~(\ref{fig3}) and find value of $\theta$ at which this difference becomes approximately zero. To show that $\theta_c$ is actually the said critical $\theta$, we plot the $G_1$ and $G_2$ decay rates against $\tau$ for values of $\theta$ less than $\theta_c$, equal to $\theta_c$, and greater than $\theta_c$ as illustrated in Fig.~(\ref{fig4}). It is clear that when $\theta = \theta_c$ (approximately $\pi/ 225$ for the case chosen), the peaks of the curves corresponding to $G_1$ and $G_2$ are at the same height above the $\tau$ axis. When $\theta < \theta_c$, the peak for $G_2$ wins, something showing that the the $x$-type behavior dominates, and when $\theta > \theta_c$, the peak for $G_1$ wins, something showing that the $x$-type behavior dominates.

\subsection*{Modified decay rates for strong and weak system-environment coupling}
\begin{figure}[h!]
\centering
\begin{subfigure}{.5\textwidth}
  \centering
  \includegraphics[width=0.95\linewidth]{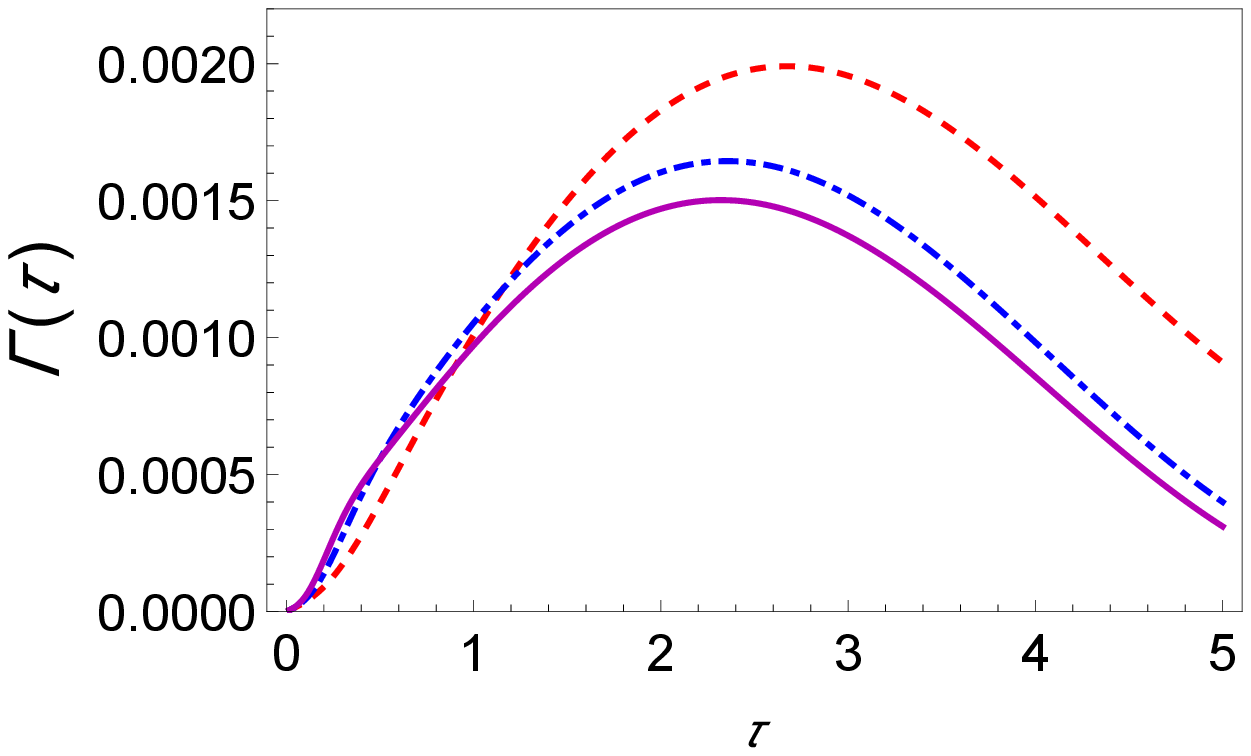}
  \caption{}
  \label{fig5a}
\end{subfigure}%
\begin{subfigure}{.5\textwidth}
  \centering
  \includegraphics[width=0.95\linewidth]{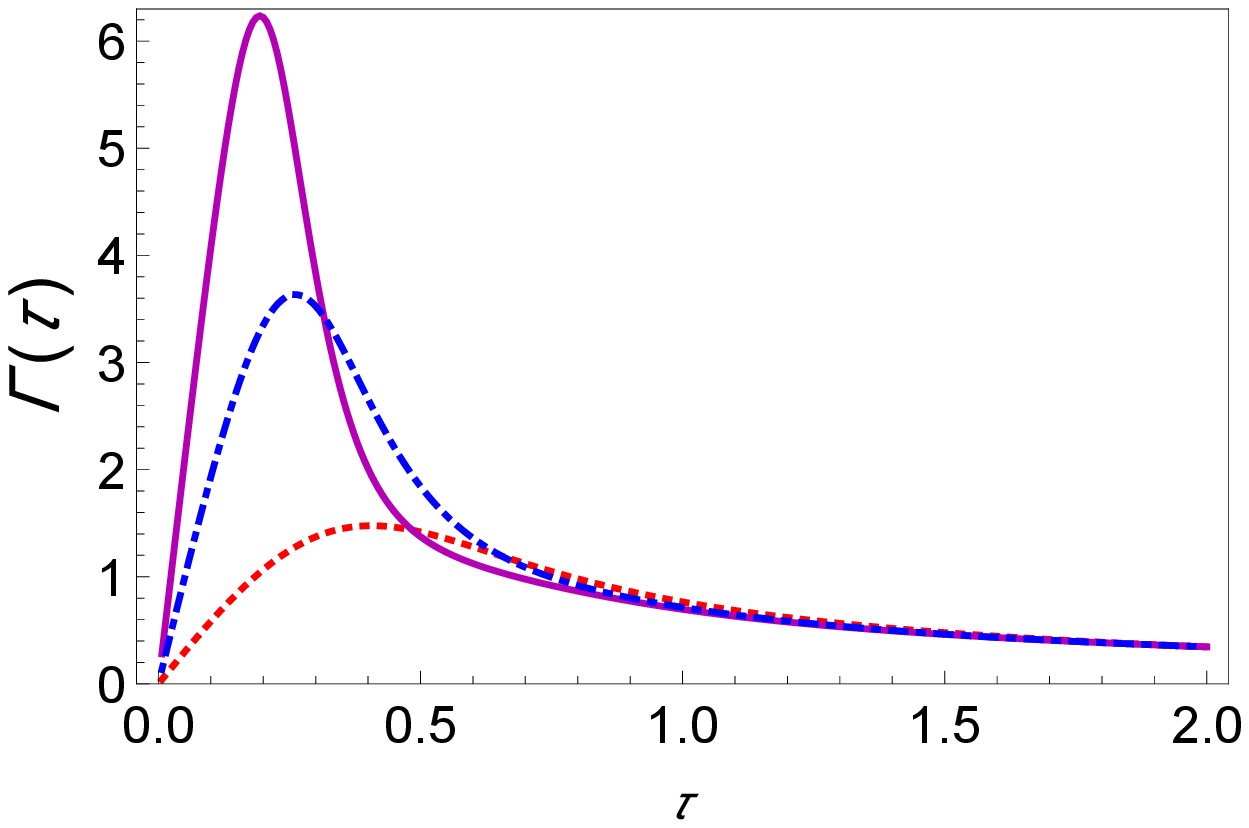}
  \caption{}
  \label{fig5b}
\end{subfigure}
\caption{\textbf{Dependence of the modified decay rate on the coupling strength.} \textbf{(a)} Graph of $\Gamma_n(\tau)$ (at zero temperature) for the initial state $\ket{0}$ as a function of $\tau$ with $G=1$ (red, dashed), $G=2$ (dot-dashed, blue), and $G=3$ (solid magenta curve). Here, we have used a super-Ohmic environment ($s=2$), with $\omega_c = 1$, $\epsilon=1$, and $\Delta = 0.05$. \textbf{(b)} Behavior of $\Gamma_n(\tau)$ (at zero temperature) for the initial state $(\ket{0}+\ket{1})/\sqrt{2}$ as a function of $\tau$ with $G=1$ (red, dashed), $G=2$ (dot-dashed, blue), and $G=3$ (solid magenta curve). Here too, we have used a super-Ohmic environment ($s=2$), with $\omega_c = 1$, $\epsilon=1$, and $\Delta = 0.05$.}
\label{fig5}
\end{figure}

In investigating the effect of changing the initial state on the QZE and the QAZE, we have used the complete Hamiltonian so far. This means that the evolution of the system state depends on the system Hamiltonian as well as the system-environment interaction. However, if we intend to study solely the effect of the dephasing reservoir on the QZE and the QAZE via its interaction with the system, we would like to remove the evolution due to the system Hamiltonian. We can do so by performing a reverse unitary time evolution due to the system Hamiltonian on the fully time-evolved density matrix as has also been done by Refs.~{\renewcommand{\citemid}{}\cite[]{Chaudhryscirep2017a, ChaudhryPRA2014zeno, Chaudhryscirep2016, MatsuzakiPRB2010}}. The survival probability becomes 
\begin{equation}
    s(\tau)=Tr_{S,B}\{P_{\psi}U_{S,I}^{\dagger}(\tau)U_{S,0}^{\dagger}(\tau)U_0(\tau) U_I(\tau) P_{\psi}\frac{e^{-\beta H}}{Z} P_{\psi} U_I^{\dagger}(\tau)U_0^{\dagger}(\tau)U_{S,0}(\tau)U_{S,I}(\tau)\},
\label{eq7}
\end{equation}
where $U_{S,0}(\tau)= e^{-iH_{S}\tau}$ and $H_{S}=\frac{\epsilon}{2}\sigma_z + \frac{\Delta}{2}\sigma_x$. 
As before, we continue to operate in the polaron frame, but it is important to note that the unitary time reversal we perform involves only the system Hamiltonian (see Methods section for details). This procedure yields the decay rate

\begin{equation}
    \Gamma_{n}(\tau) = \Gamma(\tau) + \Gamma_{\text{mod}}(\tau).
    \label{eq8}
\end{equation}

Eq.~(\ref{eq8}) shows that upon removing the system evolution, the decay rate works out to contain both the earlier found effective decay rate  and some additional terms represented by $\Gamma_{\text{mod}}(\tau)$. This simply follows from the application of perturbative approach for the second time. Now, we turn our attention to the modified decay rate expression for the system state $\ket{0}$ at zero temperature, which we find to be

\begin{equation}
    \begin{aligned}
    \begin{split}
        \Gamma_{(n)}(\tau) = \Gamma(\tau) + \frac{\Delta^2}{2}\Re{\int_{0}^{t} dt_1 \int_{0}^{t} dt_2 \bigg( e^{i\epsilon(t_1-t_2)}e^{-\Phi _{R}(t_1)} e^{-i\Phi _{I}(t_1)} \bigg) - \int_{0}^{t} dt_1 \int_{0}^{t_1} dt_2 e^{i\epsilon (t_1-t_2)}}. 
    \end{split}
    \end{aligned}
\label{eq9}
\end{equation}

The form of Eq.~(\ref{eq9}) corroborates Eq.~(\ref{eq8}) as we see that certain additional terms have emerged in the decay rate upon removal of the system evolution. However, upon simulating $\Gamma_n(\tau)$ in Fig.~(\ref{fig5a}), we observe that increasing the system-environment coupling strength decreases the decay rates as before. As such, the removal of the system evolution does not change the qualitative behavior of the decay rates in any significant way, and thus, we could confidently say that all contribution to the decay rate comes primarily from the system-environment interaction. We arrive at a similar conclusion upon plotting the decay rate corresponding to the initial state $(\ket{0} + \ket{1})/\sqrt{2}$ in Fig.~(\ref{fig5b}), that is, the qualitative behavior remains the same as before and increasing the coupling strength in fact increases the decay rates. Again, we explore the effect of varying the initial states on the modified decay rates in Fig.~(\ref{fig2b}) and find the overall analysis to be the same as before with the only contribution of the additional terms of the modified decay rate being a minor decrease in the peaks. 

In the case of the modified decay rates, we too numerically sample through decay rates corresponding to different points on the Bloch sphere to identify a transitory stage , or $\theta_c$, and find similar transitions as that of the effective case. As we move from Fig.~(\ref{fig7a}) to Fig.~(\ref{fig7c}), we observe the shift from the $x$-type behavior to the $z$-type behavior. It is interesting to note that the the critical $\theta$ also marks the value of $\theta$ that the respective magnitudes of the effective and modified decay rates invert their order at. For example, if $\theta < \theta_c$, then the modified decay rate is greater than the effective decay rate whereas it is the other way around when $\theta > \theta_c$.

\begin{figure}
    \centering
    \includegraphics[width=0.5\textwidth]{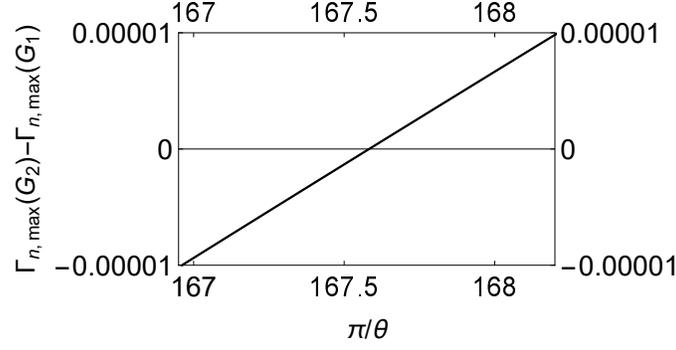}
    \caption{Graph of the difference between the maxima of the modified decay rates corresponding to $G_1$ and $G_2$, $\Gamma_{n,\text{max}}(G_2)-\Gamma_{n,\text{max}}(G_1)$ against $\pi/\theta$.}
    \label{fig6}
\end{figure}

\begin{figure}[h!]
     \centering
     \begin{subfigure}[b]{0.33\textwidth}
         \centering
         \includegraphics[width=\textwidth]{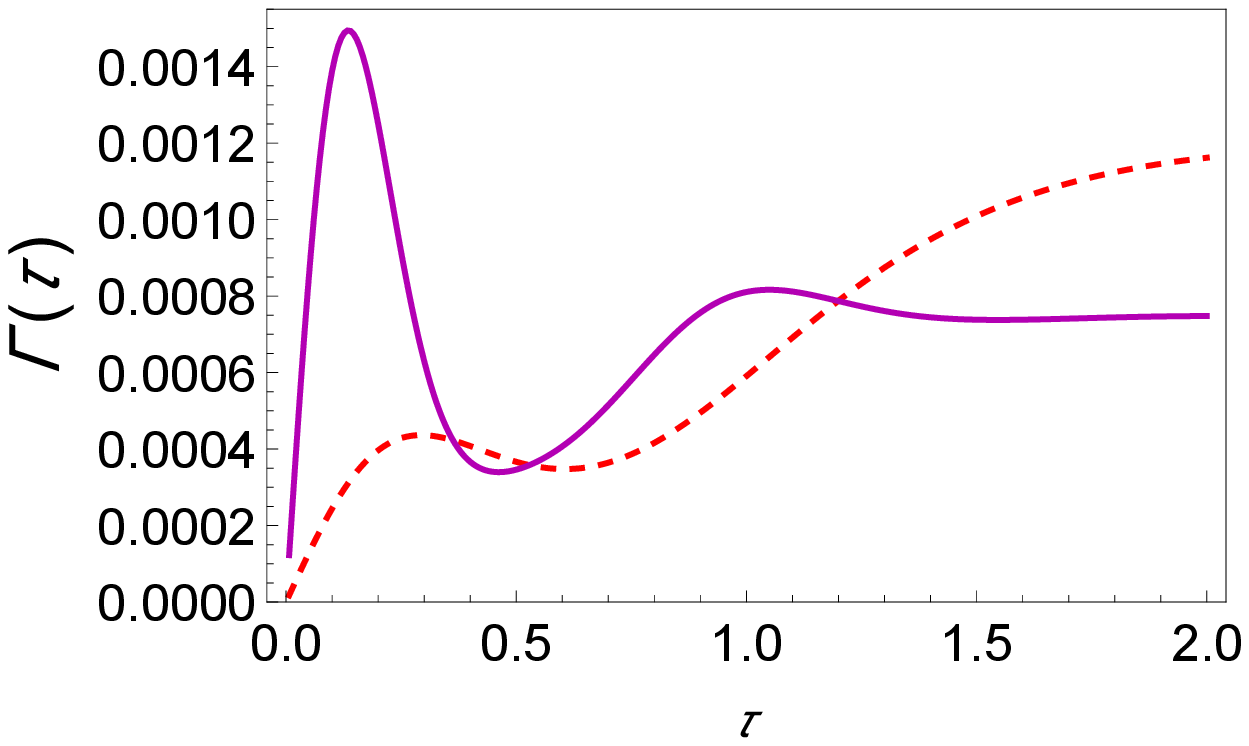}
         \caption{}
         \label{fig7a}
     \end{subfigure}
     \hfill
     \begin{subfigure}[b]{0.33\textwidth}
         \centering
         \includegraphics[width=\textwidth]{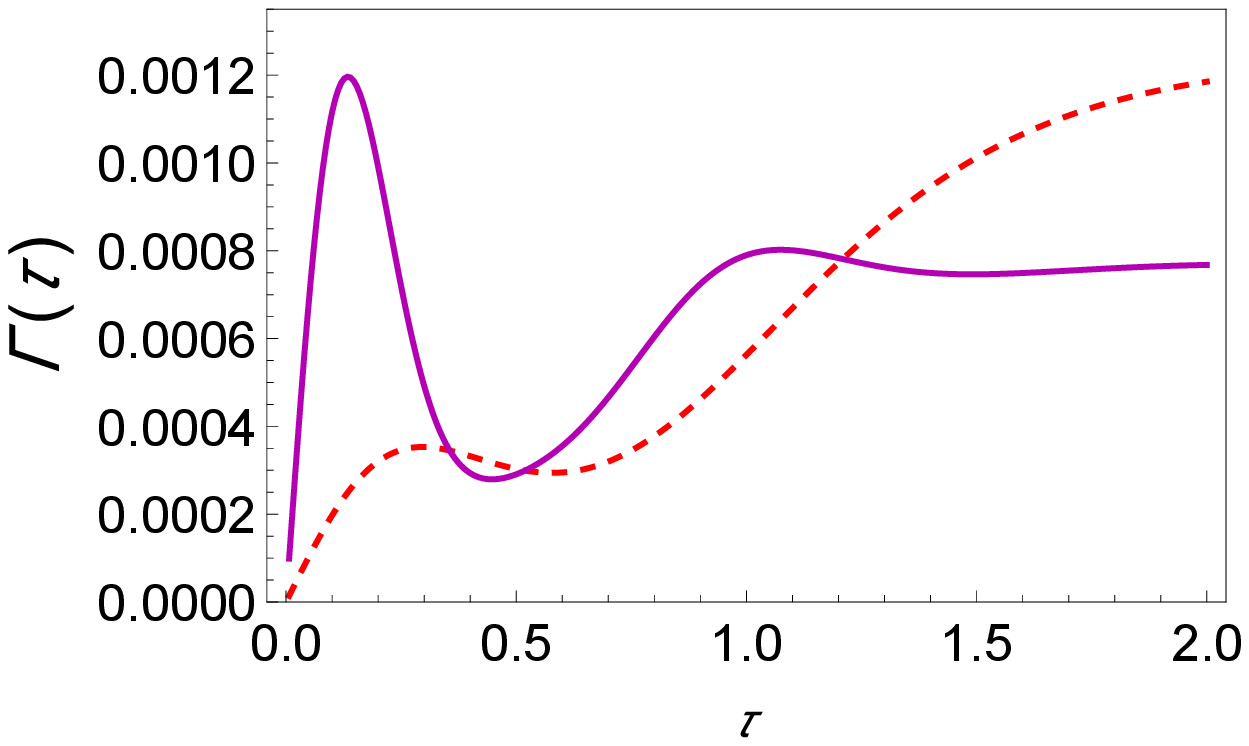}
         \caption{}
         \label{fig7b}
     \end{subfigure}
     \hfill
     \begin{subfigure}[b]{0.33\textwidth}
         \centering
         \includegraphics[width=\textwidth]{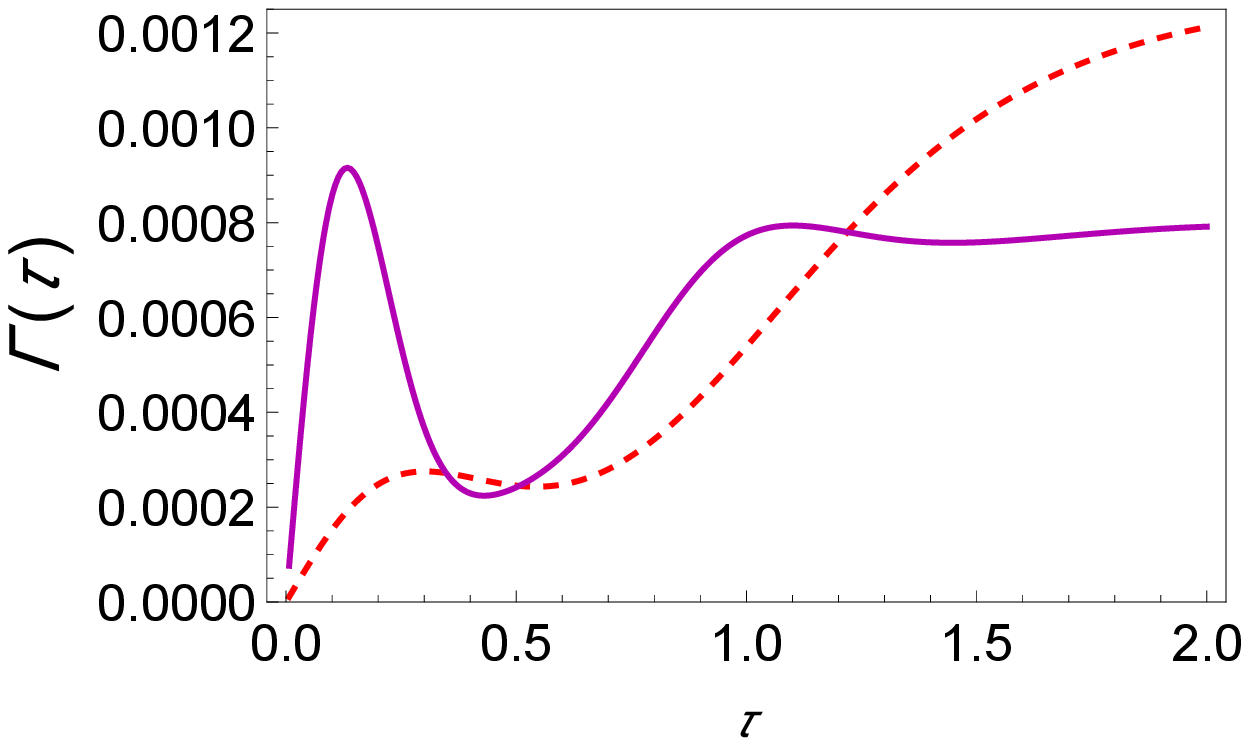}
         \caption{}
         \label{fig7c}
     \end{subfigure}
        \caption{\textbf{Transitory behavior in the modified decay rates}
        \textbf{(a)} Graph of $\Gamma_n(\tau)$ (at zero temperature) with the initial state corresponding to $\theta=\pi/150$, for $G=1$ (solid magenta curve) and $G=3$ (dashed, red curve). \textbf{(b)} Same as (a) with initial state corresponding to $\theta_c = \pi/167$, showing critical behavior. \textbf{(c)} Same as (a) except $\theta=\pi/190.$}
        \label{fig7}
\end{figure}

\section*{Discussion}

To conclude, we have extended the investigation of the QZE and the QAZE for a two-level system interacting strongly with a harmonic oscillator bath by presenting a general framework independent of the initial state chosen. We started off by transforming to the polaron frame, wherein the perturbative approach was used to make the problem tractable. From there on, we proceeded to finding the effective and modified decay rates, obtaining the latter after removing the system evolution so that the role of the environment alone may be studied. We found that the effective and modified decay rates display the same qualitative behavior, something which attests to the dominant contribution of the reservoir to the decay rates. Having set up the methodology, we continued to investigate the effect of changing the initial state on the QZE and the QAZE, which allowed us to identify the $z$-type and the $x$-type behaviors. Hence, we were able to locate critical angles about which transition between these behaviors is displayed. All these insights can be helpful for quantum control of two-level systems that are strongly interacting with a harmonic-oscillator environment.  

\section*{Methods}
\subsection*{Polaron Transformation}

Here, we present the polaron transformation for the spin-boson Hamiltonian. The transformation is given by the unitary operator $U_P= e^{-\frac{\chi\sigma_z}{2}}$ such that $H= e^{\frac{\chi\sigma_z}{2}}H_{L}e^{\frac{-\chi\sigma_z}{2}}$, where $\chi= \sum_k(\frac{2g_k^{\ast}}{\omega_k}b_k - \frac{2g_k}{\omega_k}b_k^{\dagger})$. We make use of the following identity to work this out:
\[e^{\theta X}Ye^{-\theta X} = Y+ \theta Y + \frac{\theta^2}{2!} \comm{X}{Y} \frac{\theta^3}{3!}\comm{X}{\comm{X}{Y}} +\ldots\].
Clearly, this requires that we find $\comm{\frac{\chi\sigma_z}{2}}{\sum_k \omega_k b_k^{\dagger}b_k}$, $\comm{\frac{\chi\sigma_z}{2}}{\sigma_z\sum_k(g_k^{\ast}b_k + g_k b_k^{\dagger})}$, and all their higher-order commutators. We find that $\comm{\frac{\chi\sigma_z}{2}}{\sum_k \omega_k b_k^{\dagger}b_k}=-\sigma_z\sum_k(g_k b_k^{\dagger} + g_k^{\ast}b_k)$ and that $\comm{\frac{\chi\sigma_z}{2}}{\comm{\frac{\chi\sigma_z}{2}}{\sum_k \omega_k b_k^{\dagger}b_k}}= \comm{\frac{\chi\sigma_z}{2}}{\sigma_z\sum_k(g_k^{\ast}b_k + g_k b_k^{\dagger})}=-\sum_k(\frac{\abs{g_k}^2}{\omega_k})$. Since the latter is only a constant, the higher-order commutators are zero. Moreover, since the tunneling term could be written in the form $\frac{\Delta}{2}\sigma_x = \frac{\Delta}{2}(\sigma_+ + \sigma_-)$ and $\comm{\chi\sigma_z/2}{\sigma_+}= \chi\sigma_{+}$ while $\comm{\chi\sigma_z/2}{\sigma_-}= -\chi\sigma_{-}$, we have that the tunneling term in the polaron frame is $\frac{\Delta}{2}(\sigma_+ e^{\chi} + \sigma_-e^{-\chi})$ . Upon inserting these commutators into the identity, we get

\[e^{\chi\sigma_z/2}\bigg[\frac{\epsilon}{2}\sigma_z +\frac{\Delta}{2}\sigma_x + \sum_k\omega_k b_k^{\dagger}b_k + \sigma_z\big(\sum_k g_k^{\ast}b_k^{\dagger} + g_k b_k  )\bigg]e^{-\chi\sigma_z/2} = \frac{\epsilon}{2}\sigma_z + \sum_k\omega_k b_k^{\dagger}b_k + \frac{\Delta}{2}(\sigma_+ e^{\chi} + \sigma_-e^{-\chi}),\]
which gives the polaron transformed Hamiltonian. 
\subsection*{Effective decay rate for a strongly interacting environment}

We now describe the procedure for deriving the decay rate from the survival probability stated in Eq.~(\ref{eq3}). To do so, we first work out the time-evolved density matrix of the composite system, that is, $\rho(\tau)=e^{-iHt} P_{\psi}\frac{e^{-\beta H_{0}}}{Z} P_{\psi}e^{iHt}$. Since we are in the polaron frame and we take $\Delta$ as being small, the effective system-environment interaction may be treated perturbatively. Hence, $\rho(\tau)= U_{0}(\tau)U_{I}(\tau)\rho(0)U_{I}^{\dagger}(\tau)U_{0}^{\dagger}(\tau)$, where $U_0(\tau)$ is the unitary time-evolution operator corresponding to the system Hamiltonian $H_S=\frac{\epsilon}{2}\sigma_z$ and the environment Hamiltonian $H_B =\sum_k\omega_k b_k^{\dagger}b_k$ whereas $U_I$ is the unitary evolution due to the system-environment interaction. The survival probability thus becomes
\[s(\tau)=\Tr_{S,B}\{P_{\psi}U_{0}(\tau)U_{I}(\tau)P_{\psi}\frac{e^{-\beta H_{O}}}{Z}P_{\psi}U_{I}^{\dagger}(\tau)U_{0}^{\dagger}(\tau)\}.\]

Using cyclic invariance, we absorb the system time-evolution into the projector $P_\psi$ and evolve it to $P_{\psi}(\tau)= U_{0}^{\dagger}(\tau)P_{\psi}U_{0}(\tau)$. This yields, thereby getting $P_{\psi}(\tau)=\abs{\zeta_{1}}^2 \ketbra{0}{0} + \zeta_{1}\zeta_{2}^{\ast}e^{\chi(\tau)}e^{-i\epsilon\tau}\ketbra{0}{1}+ \zeta_{2}\zeta_{1}^{\ast}e^{-\chi(\tau)}e^{-i\epsilon\tau}\ketbra{1}{0} + \abs{\zeta_{2}}^{2}\ketbra{1}{1}$. Now, we proceed to finding $U_{I}(\tau)P_{\psi}\frac{e^{-\beta H_{0}}}{Z}P_{\psi}U_{I}^{\dagger}(\tau)$. Recalling that the interaction Hamiltonian is $H_{I}=\frac{\Delta}{2}(\sigma_{+}e^{\chi} + \sigma_{-}e^{-\chi})$ in the polaron frame and writing $V_{I}(t)=e^{iH_{0} t}H_{I}(t)e^{-iH_{0} t}$, we get $V_{I}(t)= \frac{\Delta}{2}\sum_{\mu}(\widetilde{F_{\mu}}(t)\otimes\widetilde{B_{\mu}}(t))$, where $\widetilde{F_0}(t)= \sigma_{-}e^{-i\epsilon t}$, $\widetilde{F_1}( t)= \sigma_{+}e^{i\epsilon t}$, $\widetilde{B_0}(t)= e^{\chi(t)}$, and $\widetilde{B_1}(t)= e^{-\chi(t)}$. This gives $U_{I}(\tau)= 1 - i\int_0^{\tau}dt_1 V_I(t_1) + (-i)^2\int_0^{{\tau}}\int_0^{t_{1}}dt_{1}dt_{2} V_I(t_{1})V_I(t_{2})+\cdots$ Defining $A_1(\tau) = -i\int_0^{\tau}dt_1 V_I(t_1)$ and $A_2(\tau)=(-i)^2\int_0^{{\tau}}\int_0^{t_{1}}dt_{1}dt_{2} V_I(t_{1})V_I(t_{2})$, we find $\rho(\tau)$ up to the second order as:
\begin{equation}
    \rho(\tau) = \rho(0) + A_1(\tau)\rho(0) + A_2(\tau)\rho(0) + \rho(0) A_1^{\dagger}(\tau) +\rho(0) A_2^{\dagger}(\tau) + A_1(\tau)\rho(0) A_1^{\dagger}(\tau). 
    \label{eq10}
\end{equation}
It should be noted that $\rho(0)$ as given in the Results section may be written as $P_{\psi}e^{-\beta H_{0}}P_{\psi}/Z = \sum_{ijn} M_{ij}C_{ij}^{n}E_{ij}^{n}/Z$, where $i$, $j$, and $n$ could be either $0$ or $1$; $M_{ij}=\ketbra{i}{j}$; and the $C_{ij}^{n}$ and the $E_{ij}^{n}$ are given by the following tables:

\begin{table}[h]
\centering
\begin{tabular}{l | l | l| l| l}
n/ i,j & 00 & 01 & 10 & 11\\
\hline
0 & $\abs{\zeta_1}^4 e^{-\beta\epsilon/2}$ & $\abs{\zeta_1}^2\zeta_1 \zeta_2^{\ast}e^{-\beta\epsilon/2}$ & $\abs{\zeta_1}^2\zeta_2 \zeta_1^{\ast}e^{-\beta\epsilon/2}$ & $\abs{\zeta_1 \zeta_2}^2e^{-\beta\epsilon/2}$  \\
 1 & $\abs{\zeta_1 \zeta_2}^2 e^{\beta\epsilon/2}$ & $\abs{\zeta_2}^2\zeta_1 \zeta_2^{\ast}e^{\beta\epsilon/2}$ & $\abs{\zeta_2}^2\zeta_2 \zeta_1^{\ast}e^{\beta\epsilon/2}$ & $\abs{\zeta_2}^4e^{\beta\epsilon/2}$
\end{tabular}
\caption{The symbols $C^{n}_{ij}$.}
\label{columbus}
\end{table}
\begin{table}[h]
\centering
\begin{tabular}{l | l | l| l| l}
n/ i,j & 00 & 01 & 10 & 11\\
\hline
0 & $e^{-\beta H_B}$ & $e^{-\beta H_B}e^{\chi}$ & $e^{\chi}e^{-\beta H_B}$ & $e^{-\chi}e^{-\beta H_B}e^{\chi}$  \\
1 & $e^{\chi}e^{-\beta H_B}e^{-\chi}$ & $e^{\chi}e^{-\beta H_B}$ & $e^{-\beta H_B}e^{-\chi}$ & $e^{-\beta H_B}$
\end{tabular}
\caption{The symbols $E^{n}_{ij}$. Here $H_B = \sum_k \omega_k b_k^\dagger b_k$.}
\label{energy}
\end{table}
Using Tables (\ref{columbus}) and (\ref{energy}), it is easy to see that Eq.~(\ref{eq10}) may be recast as $\rho(\tau)=\sum_{i=1}^{6} T_i$, where $T_1 = \sum_{ijn}M_{ij}C_{ij}^{n}E_{ij}^{n}$, $T_2 = T_1 A_1^{\dagger}$, $T_3 = T_{1}A_{2}^{\dagger}$, $T_4 = A_{1}T_{1}$, $T_5= A_{1}T_{1}A_1^{\dagger}$, and $T_6 = A_{2}T_{1}$. Having found $\rho(\tau)$ and $P_{\psi}(\tau)$, we have $s(\tau)= \Tr_{S,B}\{{P_{\psi}(\tau)\rho(\tau)}\}$. Then, since the system-environmnet interaction is weak in the polaron frame, we may use $\Gamma(\tau)= -\frac{\ln{s(\tau)}}{\tau}$ to find the decay rate for the strongly interacting reservoir given an arbitrary initial state, $\zeta_1\ket{0}+\zeta_2\ket{1}$. The detailed expression for $\Gamma(\tau)$ may be found in the supplementary information.

\subsection*{Modified decay rate for a strongly interacting environment}
Here, we show how to work out the survival probability expressed in Eq.~(\ref{eq7}) and derive the general modified decay rate expression. In $\Tr_{S,B}\{P_{\psi}U_{S,I}^{\dagger}(\tau)U_{S,0}^{\dagger}(\tau)U_0(\tau) U_I(\tau) P_{\psi}\frac{e^{-\beta H}}{Z} P_{\psi} U_I^{\dagger}(\tau)U_0^{\dagger}(\tau)U_{S,0}(\tau)U_{S,I}(\tau)\}$, we have already worked out $U_I(\tau) P_{\psi}\frac{e^{-\beta H}}{Z} P_{\psi} U_I^{\dagger}(\tau)$ in Eq.~(\ref{eq10}). Moreover, we note that $U_{S,0}^{\dagger}(\tau)U_{0}(\tau)= e^{-iH_Bt}$. Now, we only need to work out the density matrix after the system evolution has been removed: $U_{S,I}^{\dagger}(\tau)e^{-iH_B\tau}[\rho_0 + A_1(\tau)\rho_0 + A_2(\tau)\rho_0 + \rho_0 A_1^{\dagger}(\tau) +\rho_0 A_2^{\dagger}(\tau) + A_1(\tau)\rho_0 A_1^{\dagger}(\tau)]e^{iH_B\tau}U_{S,I}(\tau)$. 
Writing $V_{S,I}(t)=e^{iH_{S,0} t}H_{I}(t)e^{-iH_{S,0} t}$, we get $V_{S,I}(t)= \frac{\Delta}{2}\sum_{\mu}(\widetilde{F_{\mu}}(t)\otimes B_{\mu})$, where $\widetilde{F_0}(t)= \sigma_{-}e^{-i\epsilon t}$, $\widetilde{F_1}( t)= \sigma_{+}e^{i\epsilon t}$, $B_0 = e^{\chi}$, and $B_1=e^{-\chi}$ as before. This leads to the interaction evolution's being given by
\begin{eqnarray}
        U_{S,I}(\tau)= 1 - i\int_0^{\tau}dt_1 V_{S,I}(t_1) + (-i)^2\int_0^{{\tau}}\int_0^{t_{1}}dt_{1}dt_{2} V_{S,I}(t_{1})V_{S,I}(t_{2})+\cdots
    \label{eq12}
\end{eqnarray}
Using $A_S^{(1)} = - i\int_0^{\tau}dt_1 V_{S,I}(t_1)$ and $A_S^{(2)} =(-i)^2\int_0^{{\tau}}\int_0^{t_{1}}dt_{1}dt_{2} V_{S,I}(t_{1})V_{S,I}(t_{2})$ now, we can conveniently write the fully time-evolved density matrix with the system evolution removed as $\rho(\tau) = (1 + A_S^{(1)\dagger}+ A_S^{(2)\dagger})(e^{-iH_B t}\sum_{i=1}^{N=6}T_{i}e^{iNt})(1 + A_S^{(1)}+ A_S^{(2)})$. We work this out to second order, apply the projection operator $P_{\psi}$, and find the system and bath traces the same way as before. It then is straightforward to arrive at the survival probability that the modified decay rate could be found from. From Eq.~(\ref{eq12}), it is easily verified that the modified decay rate is simply the sum of the effective decay rate found earlier and some additional terms that we denote with $\Gamma_{\text{mod}}$. As such, we may write $\Gamma_{n}(\tau) = \Gamma(\tau) + \Gamma_{\text{mod}}(\tau)$.

\setcounter{equation}{0}

\section*{Supplementary Material: The quantum Zeno and anti-Zeno effects in the strong coupling regime}

Continuing with the notation described in the Methods section, the following works out to be the effective decay rate for an arbitrarily initialized state
\begin{equation}
    \begin{aligned}
    \begin{split}
        \Gamma(\tau)=& \frac{1}{\tau}\bigg[1 -\sum_{n}\Tr_{B}\bigg\{\frac{1}{Z}\bigg\{\abs{\zeta_1}^2 C_{00}^{n}E_{00}^{n}
        +  \zeta_{1}\zeta_{2}^{*}e^{i\epsilon t}C_{10}^{n}e^{\chi(t)}E_{10}^{n} + \zeta_{1}\zeta_{2}^{*}e^{i\epsilon t}C_{10}^{n}e^{-\chi(t)}E_{10}^{n} + \abs{\zeta_2}^2 C_{11}^{n}E_{11}^{n}\\
        &+ 2\Re{ 
        i\frac{\Delta}{2}\int_{0}^{t}dt_1 \abs{\zeta_1}^2 e^{i\epsilon t_1}C_{01}^{n}E_{01}^{n}\widetilde{B_{0}}
        +\abs{\zeta_2}^2 e^{i\epsilon t_1}C_{10}^{n}E_{10}^{n}\widetilde{B_{1}}
        +\zeta_1\zeta_2^{\ast}e^{\chi(t)}e^{i\epsilon(t-t_1)}
        C_{11}^{n}E_{11}^{n}\widetilde{B_{0}}
        +\zeta_2\zeta_1^{\ast}e^{-\chi(t)}e^{-i\epsilon(t-t_1)}
        C_{00}^{n}E_{00}^{n}\widetilde{B_{1}}}\\
        &-2\Re{
        \frac{\Delta^2}{4}\int_{0}^{\tau} dt_1\int_{0}^{t_1} dt_2
        \abs{\zeta_1}^2 \Omega^{01}C_{00}^{n}E_{00}^{n}\Bar{B_1}\widetilde{B_0}
        +\abs{\zeta_2}^2\Omega^{10}C_{11}^{n}E_{11}^{n}\Bar{B_0}\widetilde{B_1}
        +\zeta_1\zeta_2^{\ast}e^{\chi(t)}e^{i\epsilon t}\Omega^{01}C_{10}^{n}E_{10}^{n}\Bar{B_1}\widetilde{B_0}}\\
        &+\zeta_2\zeta_1^{\ast}e^{-\chi(t)}e^{-i\epsilon}t\Omega^{10}C_{11}^{n}E_{11}^{n}\Bar{B_0}\widetilde{B_{1}}\\ 
        &+\frac{\Delta^2}{4} \int_{0}^{\tau}dt_1\int_{0}^{\tau}dt_2  \abs{\zeta_1}^2\Omega^{10}C_{11}^{n}\widetilde{B_1}E_{11}^{n}\Bar{B_0}
        +\zeta_1\zeta_2^{\ast}e^{i\epsilon t}e^{\chi(t)}\Omega^{00}C_{01}^{n}\widetilde{B_0}E_{01}^{n}\Bar{B_0}\\
        &+\zeta_2\zeta_1^{\ast}e^{-i\epsilon t}e^{-\chi(t)}\Omega^{11}C_{10}^{n}\widetilde{B_1}E_{10}^{n}\Bar{B_1}
        +\abs{\zeta_2}^2\Omega^{01}C_{00}^{n}\widetilde{B_0}E_{00}^{n}\Bar{B_1}
        \bigg\}\bigg],
    \end{split}
    \end{aligned}
\label{A1}
\end{equation}
where $\Omega^{00} = e^{-i\epsilon(t_1 + t_2)}$, $\Omega^{01} = e^{-i\epsilon(t_1 - t_2)}$, $\Omega^{10} = (\Omega^{01})^{\dagger}$, and $\Omega^{11}= (\Omega^{00})^{\dagger}$. $\widetilde{B_0}= e^{-\chi(t_1)}$ and $\widetilde{B_1}= e^{\chi(t_1)}$. From Eq.~(\ref{A1}), it is easy to see the bath traces that would yield the correlation functions. To find the bath traces, we use Bloch's identity. As an example, we work out 
\[\bigg<e^{-\chi(\tau)}e^{\chi(t_1)}e^{\chi}e^{\chi(t_2)}\bigg>_B =
e^{-\comm{\chi(\tau)}{\chi(t_1)}/2} e^{-\comm{\chi(t_1)}{\chi}/2} e^{\comm{\chi(t_1)}{\chi}/2}e^{-\comm{\chi(t)}{\chi(t_2)}/2}e^{\comm{\chi(t_1)}{\chi(t_2)}/2}e^{\comm{\chi}{\chi(t_2)}/2}
e^{\bigg<(-\chi(\tau)+\chi(t_1)+\chi+\chi(t_2))^2/2\bigg>_{B}}.\]
This works out to be $W e^{-i\Phi_{I}(t_2)}e^{-i\Phi_{I}(t_1)}e^{i\Phi_{I(\tau)}}$, where $W$ is defined as
\[W=e^{-2\Phi_{R2}}e^{-\Phi_{R1}(t_{2}-t_{1})}e^{\Phi_{R1}(t_{2}-t)}e^{\Phi_{R1}(t_{1}-t)}e^{\Phi _{R1}(t_{2})}e^{\Phi _{R1}(t_{1})}e^{-\Phi_{R1}(t)}e^{i\Phi_{I}(t_{2}-t_{1})}e^{-i\Phi_{I}(t_{2}-t)}e^{i\Phi_{I}(t_{1}-t)}.\]
Here $\Phi_{C} = \Phi_{R} - i\Phi_{I}$, $\Phi_{R}= 4\int_{0}^{\infty}d\omega J(\omega)\frac{1-\cos{(\omega \tau)}}{\omega^2}\coth({\frac{\beta\omega}{2}})$, $\Phi_{R1}(t)=4\int_{0}^{\infty}d\omega J(\omega)\frac{\cos{(\omega \tau)}}{\omega^2}\coth({\frac{\beta\omega}{2}})$,\\ 
$\Phi _{R2}(t)=4\int_{0}^{\infty}d\omega J(\omega)(\coth({\frac{\beta\omega}{2}})/\omega^2)$, and $\Phi _{I}=4\int_{0}^{\infty}d\omega J(\omega) \frac{ \sin{(\omega\tau)}}{\omega^2}$, where the environment spectral densities have been introduced as $\sum_{k}\abs{g_k}^2(\cdots) \rightarrow \int_{0}^{\infty}d\omega J(\omega) (\cdots)$. We also note
\[W'=e^{-2\Phi_{R2}}e^{\Phi_{R1}(t_{2}-t_{1})}e^{-\Phi_{R1}(t_{2}-t)}e^{\Phi_{R1}(t_{1}-t)}e^{\Phi_{R1}(t_{2})}e^{-\Phi _{R1}(t_{1})}e^{\Phi_{R1}(t)}e^{-i\Phi_{I}(t_{2}-t_{1})}e^{i\Phi_{I}(t_{2}-t)}e^{-i\Phi_{I}(t_{1}-t)},\]
as this appears in other bath traces. Having found all the traces, we can now write our effective decay rate expression:
\begin{equation}
    \begin{aligned}
    \begin{split}
    \Gamma^{(0)}(\tau)=& \frac{1}{\tau}\bigg[1- 
    \frac{1}{Z_s}\bigg[
    \abs{\zeta_{1}}^6e^{-\beta\epsilon/2}
    + \abs{\zeta_{1}}^4\abs{\zeta_{2}}^2 e^{\beta\epsilon/2}
    + \abs{\zeta_{1}}^2\abs{\zeta_{2}}^4 e^{-\beta\epsilon/2}
    + \abs{\zeta_{2}}^6 e^{\beta\epsilon/2}\\
    &+ 2\Re\bigg\{\abs{\zeta_{1}}^2\abs{\zeta_{2}}^4 e^{i\epsilon\tau}e^{\beta\epsilon/2}e^{-\Phi_{C}(\tau)}
    +\abs{\zeta_{1}}^4\abs{\zeta_{2}}^2 e^{i\epsilon\tau} e^{-\beta\epsilon/2}e^{-\Phi_{C}^{\ast}(\tau)}\bigg\}\\
    &+2\Re\bigg\{i\frac{\Delta}{2}\int_{0}^{\tau}dt_1 \bigg(\abs{\zeta_{1}}^4\zeta_{1}\zeta_{2}^{\ast} e^{-i\epsilon t_1}e^{-\beta\epsilon/2}e^{-\Phi_{C}(t_1)}+\abs{\zeta_{1}\zeta_{2}}^2\zeta_{1}\zeta_{2}^{\ast} e^{-i\epsilon t_1}e^{\beta\epsilon/2}e^{-\Phi_{C}^{\ast}(t_1)}\\
    &+\abs{\zeta_{1}\zeta_{2}}^2\zeta_{2}\zeta_{1}^{\ast} e^{i\epsilon t_1}e^{-\beta\epsilon/2}e^{-\Phi_{C}^{\ast}(t_1)}+ \abs{\zeta_{2}}^4\zeta_{2}\zeta_{1}^{\ast} e^{i\epsilon t_1}e^{\beta\epsilon/2}e^{-\Phi_{C}(t_1)}\\
    &+\abs{\zeta_{2}}^4\zeta_{1}\zeta_{2}^{\ast}e^{i\epsilon\tau}e^{-i\epsilon t_1}e^{\beta\epsilon/2}e^{-\Phi_{C}^{\ast}(t_1-\tau)}+\abs{\zeta_{1}\zeta_{2}}^2\zeta_{1}\zeta_{2}^{\ast}e^{i\epsilon\tau} e^{-i\epsilon t_1}e^{-\beta\epsilon/2}e^{-\Phi_{C}^{\ast}(t_1-\tau)}e^{-2i\Phi _{I}(\tau)}e^{2i\Phi_{I}(t_1)}\\
    &+\abs{\zeta_{1}}^4\zeta_{2}\zeta_{1}^{\ast}e^{-i\epsilon\tau} e^{i\epsilon t_1}e^{-\beta\epsilon/2}e^{-\Phi_{C}^{\ast}(t_1-\tau)}+\abs{\zeta_{1}\zeta_{2}}^2\zeta_{2}\zeta_{1}^{\ast}e^{-i\epsilon\tau} e^{i\epsilon t_1}e^{\beta\epsilon/2}e^{-\Phi_{C}^{\ast}(t_1-\tau)}e^{-2i\Phi_{I}(\tau)}e^{2i\Phi_{I}(t_1)}\bigg)\bigg\}\\
    &-2\Re\bigg\{\frac{\Delta^2}{4}\int_{0}^{\tau}dt_1\int_{0}^{t_1}dt_2 \bigg( 
    \abs{\zeta_{1}}^6 e^{-i\epsilon t_1}e^{i\epsilon t_2}e^{-\beta\epsilon/2}e^{-\Phi_{C}^{\ast}(t_2-t_1)}\\
    &+\abs{\zeta_{1}}^4\abs{\zeta_{2}}^2 e^{-i\epsilon t_1}e^{i\epsilon t_2}e^{\beta\epsilon/2}e^{-\Phi_{C}^{\ast}(t_2-t_1)}e^{2i\Phi _{I}(t_2)}e^{-2i\Phi_{I}(t_1)}\\
    &+\abs{\zeta_{1}}^2\abs{\zeta_{1}\zeta_{2}}^2 e^{i\epsilon t_1}e^{-i\epsilon t_2}e^{-\beta\epsilon/2}e^{-\Phi_{C}^{\ast}(t_2-t_1)}e^{2i\Phi_{I(t_2)}}e^{-2i\Phi_{I(t_1)}}\\
    &+\abs{\zeta_{2}}^6 e^{i\epsilon t_1}e^{-i\epsilon t_2}e^{\beta\epsilon/2}e^{-\Phi_{C}^{\ast}(t_2-t_1)}+
    \abs{\zeta_{1}\zeta_{2}}^2\abs{\zeta_{1}}^2 e^{i\epsilon\tau}e^{-i\epsilon t_1}e^{i\epsilon t_2}e^{-\beta\epsilon/2}W e^{-i\Phi_{I}(t_2)}e^{i\Phi _{I}(t_1)}e^{-i\Phi_{I}(\tau)}\\
    &+\abs{\zeta_{1}\zeta_{2}}^2\abs{\zeta_{2}}^2 e^{i\epsilon\tau}e^{-i\epsilon t_1}e^{i\epsilon t_2}e^{\beta\epsilon/2}W' e^{i\Phi _{I}(t_2)}e^{-i\Phi _{I}(t_1)}e^{i\Phi_{I}(\tau)}
    +\abs{\zeta_{1}\zeta_{2}}^2\abs{\zeta_{1}}^2 e^{-i\epsilon\tau}e^{i\epsilon t_1}e^{-i\epsilon t_2}e^{-\beta\epsilon/2}W'e^{i\Phi_{I}(t_2)}e^{-i\Phi_{I}(t_1)}e^{i\Phi_{I}(\tau)}\\
    &+\abs{\zeta_{1}\zeta_{2}}^2\abs{\zeta_{1}}^2 e^{-i\epsilon\tau}e^{i\epsilon t_1}e^{-i\epsilon t_2}e^{\beta\epsilon/2}W'e^{-i\Phi_{I}(t_2)}e^{i\Phi _{I}(t_1)}e^{-i\Phi_{I}(\tau)}\bigg)\bigg\}\\ 
    &+\frac{\Delta^2}{4}\int_{0}^{\tau}dt_1\int_{0}^{\tau}dt_2\bigg(\abs{\zeta_{1}\zeta_{2}}^2\abs{\zeta_{1}}^2 e^{i\epsilon t_1}e^{-i\epsilon t_2}e^{-\beta\epsilon/2}
    e^{-\Phi_{C}^{\ast}(t_2-t_1)}e^{2i\Phi_{I}(t_2)}e^{-2i\Phi_{I}(t_1)}\\
    &+\abs{\zeta_{1}\zeta_{2}}^2\abs{\zeta_{2}}^2 e^{i\epsilon t_1}e^{-i\epsilon t_2}e^{\beta\epsilon/2}e^{-\Phi_{C}^{\ast}(t_2-t_1)}+\abs{\zeta_{1}\zeta_{2}}^2\abs{\zeta_{1}}^2 e^{-i\epsilon t_1}e^{i\epsilon t_2}e^{-\beta\epsilon/2}
    e^{-\Phi_{C}^{\ast}(t_2-t_1)}\\
    &+\abs{\zeta_{1}\zeta_{2}}^2\abs{\zeta_{2}}^2 e^{-i\epsilon t_1}e^{i\epsilon t_2}e^{\beta\epsilon/2}
    e^{-\Phi_{C}^{\ast}(t_2-t_1)}e^{2i\Phi_{I}(t_2)}e^{-2i\Phi_{I}(t_1)}\\
    &+\abs{\zeta_{1}}^2\zeta_{1}^2\zeta_{2}^{\ast2}e^{i\epsilon\tau} e^{-i\epsilon t_1}e^{-i\epsilon t_2}e^{-\beta\epsilon/2}W e^{i\Phi _{I(t_2)}}e^{i\Phi_{I}(t_1)}e^{-i\Phi_{I}(\tau)}
    +\abs{\zeta_{2}}^2\zeta_{1}^2 \zeta_{2}^{\ast2}e^{i\epsilon\tau} e^{-i\epsilon t_1}e^{-i\epsilon t_2}e^{\beta\epsilon/2}W e^{-i\Phi _{I}(t_2)}e^{-i\Phi_{I}(t_1)}e^{i\Phi_{I}(\tau)}\\
    &+\abs{\zeta_{1}}^2\zeta_{2}^2 \zeta_{1}^{\ast2}e^{-i\epsilon\tau} e^{i\epsilon t_1}e^{i\epsilon t_2}e^{-\beta\epsilon/2}W e^{-i\Phi _{I}(t_2)}e^{-i\Phi_{I}(t_1)}e^{i\Phi_{I}(\tau)}
    +\abs{\zeta_{2}}^2\zeta_{2}^2\zeta_{1}^{\ast2}e^{-i\epsilon\tau} e^{i\epsilon t_1}e^{i\epsilon t_2}e^{\beta\epsilon/2}W e^{i\Phi _{I}(t_2)}e^{i\Phi_{I}(t_1)}e^{i\Phi_{I}(\tau)}
    \bigg)\bigg]\bigg]
    \end{split}
    \end{aligned}
\label{eq_strong_normal}
\end{equation}
We model the spectral density as $J(\omega)= G\omega^{s}\omega_{c}^{1-s}e^{-\omega/\omega_c}$ where $G$ is a dimensionless parameter characterizing the strength of the system-environment coupling, $\omega_c$ is the cut off frequency, and $s$ is the Ohmicity parameter. Setting $s=2$ which corresponds to the super-Ohmic spectral density yields $\Phi_{R}=G(4-\frac{4}{1+\omega_c^2 t^2})$, $\Phi_{R1}=\frac{4G}{1 + \omega_c^2 t^2}$, $\Phi_{R2}= 4G$, and $\Phi_{I}= \frac{4Gt}{(\omega_c(\frac{1}{\omega_c^2}+t^2))}$.

\end{document}